\documentclass{Interspeech2024}
\usepackage{booktabs}
\usepackage{multirow,cite}
\usepackage[caption=false]{subfig}

\interspeechcameraready

\title{Toward Fully-End-to-End Listened Speech Decoding from EEG Signals}

\name[]{Jihwan}{Lee}
\name[]{Aditya}{Kommineni}
\name[]{Tiantian}{Feng}
\name[]{Kleanthis}{Avramidis}
\name[]{Xuan}{Shi}
\name[]{Sudarsana Reddy}{Kadiri}
\name[]{Shrikanth}{Narayanan}

\address{
  University of Southern California, USA}
\email{jihwan@usc.edu, akommine@usc.edu, tiantiaf@usc.edu, avramidi@usc.edu, xuanshi@usc.edu, skadiri@usc.edu, shri@usc.edu}


\begin{document}

\maketitle

\begin{abstract}

Speech decoding from EEG signals is a challenging task, where brain activity is modeled to estimate salient characteristics of acoustic stimuli.
We propose FESDE, a novel framework for \textbf{F}ully-\textbf{E}nd-to-end \textbf{S}peech \textbf{D}ecoding from \textbf{E}EG signals.
Our approach aims to directly reconstruct listened speech waveforms given EEG signals, where no intermediate acoustic feature processing step is required.
The proposed method consists of an EEG module and a speech module along with a connector. The EEG module learns to better represent EEG signals, while the speech module generates speech waveforms from model representations. The connector learns to bridge the distributions of the latent spaces of EEG and speech. The proposed framework is both simple and efficient, by allowing single-step inference, and outperforms prior works on objective metrics. A fine-grained phoneme analysis is conducted to unveil model characteristics of speech decoding.
The source code is available here: \url{github.com/lee-jhwn/fesde}. 


\end{abstract}

\section{Introduction}

Brain-computer interfaces (BCI), particularly those targeting decoding listened or articulated speech from neural (brain activity) signals such as electroencephalography (EEG) and electrocorticography (ECoG), hold promise for improving life quality and rehabilitation outcomes for patients with communication disorders.
Speech decoding technologies from brain activity also offer possibilities for seamless, immersive, and interactive entertainment, where command and control are coordinated without compromising sensitive information accessible to others.
For instance, a research group recently successfully prototypes a digital avatar of a paralyzed patient relying on understanding and interpretation of ECoG signals \cite{metzger2023high} enabling the patient to communicate more freely.

Several approaches have been proposed to decode speech or text from neural activities such as EEG or ECoG signals \cite{metzger2023high, puffay2023relating, lopez2022state,willett2023high, li2023neural2speech, meta_paper}.
For example, many efforts have proposed the use of convolution neural networks (CNN) to decode speech from EEG \cite{vlaai, xu2024convconcatnet} and ECoG signals \cite{Angrick_2019} especially to reconstruct speech signals or representations, like mel-spectrogram, from brain activity signals.
Furthermore, several methods to decode imagined speech from neural activities have also been proposed by \cite{imaginedspeech23,proix2022imagined}.

Decoding speech from EEG signals rather than ECoG signals is more practical and economical in real world scenarios.
Acquiring ECoG signals is challenging and limited to certain types of patients, as a surgical intervention is necessary to implant the sensing electrodes.
On the other hand, obtaining EEG signals only requires participants to wear a cap-like device.
Hence, speech decoding frameworks based on EEG signals can easily be applied to larger group of users, while ECoG based approaches are primarily focused on patients with specific clinical conditions such as epilepsy or certain movement disorders.



However, modeling and decoding information from EEG recordings has been a significant signal processing challenge, due to the low signal-to-noise ratio (SNR) that is inherent in EEG measurements and sparsity of measurements. EEG devices attempt to measure electrical activity in the brain through non-invasive electrodes, spatially arranged on the scalp. Hence, the sensed signals are weakened, with low spatial resolution, and contaminated by various biological, environmental noises, such as muscle activity, cardiac activity, eye movements, power lines, or nearby electronic devices~\cite{uriguen2015eeg, michel2019eeg}.
Various approaches have been proposed to deal with these sources of variability, ranging from sophisticated pre-processing and decomposition~\cite{uriguen2015eeg} algorithms to representation learning frameworks that are trained in a label-agnostic, self-supervised regime~\cite{kostas2021bendr,kommineni2024knowledgeguided}.


Decoding speech waveforms {\em directly} from EEG signals has numerous advantages when compared to multi-step approaches that usually incorporate signal-to-text estimation as an intermediate step.
Speech contains rich information such as prosody including rhythm, intonation, emotion, and speaker identity that are lost in the typical written lexical form.
Hence, a pipelined approach to processing EEG signals that first goes through text decoding, followed by text-to-speech (TTS), faces challenges in decoding high-fidelity speech from the loss of important prosodic and other paralinguistic information.

Recent development in TTS technology may facilitate direct speech decoding from EEG signals.
Fully-end-to-end TTS systems, such as VITS \cite{vits}, are TTS systems that directly synthesize speech in waveform, rather than intermediate acoustic features like mel-spectrograms or mel-frequency cepstral coefficients (MFCC) \cite{tan2021survey}.
Fully-end-to-end TTS systems have numerous advantages over other TTS systems that require any additional intermediate acoustic feature mapping step.
As the intermediate step is skipped, fully-end-to-end TTS systems are faster, simpler, and less vulnerable to error accumulation, compared to TTS systems that require an extra step of converting acoustic features to final waveforms.
With such recent developments in the TTS domain, we believe the intermediate acoustic feature extraction step can be omitted. 


In this paper, we propose \textbf{FESDE}, a novel framework for \textbf{F}ully-\textbf{E}nd-to-end \textbf{S}peech \textbf{D}ecoding from \textbf{E}EG signals.
Our proposed framework directly generates waveform of listened speech given EEG signals, without any additional intermediate acoustic feature mapping step, such as to mel-spectrogram. To the best of our knowledge, this is the first framework that directly generates speech waveform given respective EEG signals. FESDE consists of three parts: the EEG module, the speech module, and the connector. The EEG module learns to provide a better representation of EEG signals. The speech module aims to generate the speech waveform from the speech embeddings. The connector learns to map and convert the distribution of EEG embedding to speech embedding. Figure~\ref{fig:overall-archi} illustrates the overall structure of the proposed framework. During inference, only the EEG encoder and the speech decoder are utilized, along with the connector.

The proposed framework has various advantages over previous approaches. It enables single-step inference and does not require any additional pipelines such as a vocoder. Hence, it is faster and more straightforward compared to previous multi-step approaches, while outperforming the previous approach in terms of objective metrics.

The main contributions of the paper are as follows:
\begin{itemize}
    \item We propose a fully-end-to-end speech decoding framework for EEG signals, by incorporating EEG module and speech module, reducing multiple steps into one. To our best knowledge, this is the first attempt to directly reconstruct listened speech waveform from EEG signals.
    \item The proposed framework allows single-step inference, which is simpler and faster, while also outperforming the previous approaches in objective measures.
    \item We conduct phoneme analysis and present the characteristics of phonemes that are easy or challenging to decode by this approach.
\end{itemize}

\section{Proposed Method}
\subsection{Model Architecture}

\begin{figure}[ht]
  \centering
  \includegraphics[width=\linewidth]{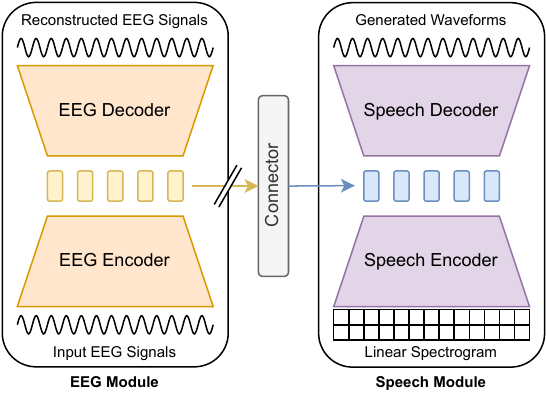}
  \caption{The overall schematic of the proposed method. The EEG module is trained to produce descriptive representations of EEG signals. The speech module aims to generate speech waveform from the speech embeddings. The connector converts the distribution of the EEG embedding into the speech embedding. During inference, only the EEG encoder and the speech decoder are utilized, along with the connector.}
  \label{fig:overall-archi}
\end{figure}

As shown in Figure~\ref{fig:overall-archi}, the proposed framework consists of three parts: the EEG module, the speech module, and the connector.
The EEG and speech signals are handled by their respective modules. The connector bridges the two intermediate embeddings from EEG and speech.
During inference, only the EEG encoder, the connector, and the speech decoder are used.
For detailed implementation, refer to the source code.

\subsubsection{EEG Module}

The EEG module is based on \cite{kommineni2024knowledgeguided}, a self-supervised learning framework for EEG signal representation.
The EEG module consists of an encoder, that learns to encode input EEG signals into intermediate EEG embeddings, and a decoder, that learns to reconstruct EEG signals from those embeddings.
The EEG encoder consists of convolution blocks and Structured State Space Sequence (S4) layers \cite{s4_paper}.
Each convolution block is composed of a 1D convolution layer, a dropout layer, layer normalization, and GELU activation.
Each S4 layer consists of an S4 kernel estimator, gated linear unit (GLU) activation, dropout, and layer normalization.
The S4 layers show superior performance in encoding long range temporal dependencies which make them suitable for encoding EEG signal \cite{kommineni2024knowledgeguided}.
The EEG decoder is composed of deconvolution blocks, where each block contains a 1D transpose convolution layer along with a dropout layer, layer normalization, and GELU activation.

\subsubsection{Speech Module}
We base the speech module on VITS \cite{vits}, one of the well performing fully-end-to-end TTS frameworks.
The speech module consists of two parts: the speech encoder and the speech decoder. The speech encoder takes linear spectrograms as input and outputs the intermediate speech embeddings. It is composed of non-causal WaveNet \cite{vandenoord16wavenet} residual blocks, as in \cite{vits, waveglow, glowtts}, and a projection layer that outputs the mean and the variance of the distribution of the speech embeddings.
We adopt HiFi-GAN V1 \cite{hifigan} as the speech decoder, as in \cite{vits}.

\subsubsection{Connector}
The connector consists of two parts: the prenet and the flow. As in \cite{vits}, the prenet consists of the transformer encoder \cite{transformer} and a linear projection layer. It takes the intermediate EEG embeddings as input and outputs mean and variance of the distribution. The flow is identical to the normalizing flow network in \cite{vits}, with a stack of affine coupling layers \cite{dinh2017density}. The flow works as an invertible function between the two distributions of the EEG and speech embeddings. There is a gradient stop between the intermediate EEG embeddings and the connector. We empirically observed that the gradient stop stabilizes the training of the EEG module, as it is not affected by the performance of the speech module in early stages of training.

\subsection{Training Objectives}
Employing the approach in \cite{kommineni2024knowledgeguided}, the cosine similarity loss is used for training the EEG module, as in Eq.~(\ref{equation:loss-cossim}):
\begin{align}
L_{\text{EEG}}(x, \hat{x}) = 1 - \frac{1}{N_{\text{ch}}}\sum_{i=1}^{N_{\text{ch}}}\frac{x_i^T \cdot \hat{x_i}}{\Vert x_i \Vert \Vert \hat{x_i}\Vert}
  \label{equation:loss-cossim}
\end{align}
where $N_{\text{ch}}$ is the number of EEG channels, and $x$ and $\hat{x}$ represent the input and the reconstructed EEG signals, respectively.

We adopt training objectives for the speech module from \cite{vits, hifigan}. The reconstruction loss $L_{\text{mel}}$ is defined as the L1 loss between the mel-spectrograms of generated and actual speech waveform. 
The KL-divergence loss $L_{\text{KL}}$ helps map the distributions of the intermediate embeddings of speech and EEG to one another, as in Eq.~(\ref{equation:loss-kld}).
\begin{align}
L_{\text{KL}}(p, q) = \log{q(z|y)} - \log{p(z|x)}
  \label{equation:loss-kld}
\end{align}
where $y$ is the input speech and $z$ is the intermediate speech embedding.
Hence, $q(z|y)$ and $p(z|x)$ represent the distribution of the intermediate speech embeddings given $y$ (speech) and $x$ (EEG signals), respectively.
The total loss for the speech module $L_{\text{speech}}$ is as follows:
\begin{align}
\vspace{-0.5cm}
L_{\text{speech}} = L_{\text{mel}} + L_{\text{KL}} + L_{\text{GAN}}
  \label{equation:loss-total}
  \vspace{-0.5cm}
\end{align}
where $L_{\text{GAN}}$ consists of the adversarial loss and the feature matching loss from HiFi-GAN V1 \cite{hifigan}.
Note that gradient stop is applied between the EEG module and the connector, the training of the EEG module is not affected by the speech module.


\begin{table*}[ht]
\footnotesize
  \caption{MCD (dB) and Mel-Corr (\%) of each configuration, with 95\% confidence intervals.}
  \label{tab:mcd-corr}
  \vspace{-0.2cm}
  \centering
  \begin{tabular}{cc|ccc|ccc}
    \toprule
    \multicolumn{2}{c|}{\multirow{2}{*}{\textbf{Model}}} & \multicolumn{3}{c|}{\textbf{MCD (dB) $\downarrow$}} & \multicolumn{3}{c}{\textbf{Mel-Corr (\%) $\uparrow$}}\\
     \multicolumn{2}{c|}{}&unseen audio&unseen subject&unseen both&unseen audio&unseen subject&unseen both\\
    \toprule
    \multicolumn{2}{c|}{\textsc{baseline} \cite{vlaai}} & $13.00 \pm 0.12$ & $13.20 \pm 0.12$ & $12.82 \pm 0.39$ & $13.27 \pm 0.66$ & $12.66 \pm 0.74$ & $13.07 \pm 2.06$ \\
    \midrule
    \multirow{5}{*}{\textsc{Ours}} & \textsc{vanilla} & $11.67 \pm 0.12$ & $11.72 \pm 0.11$ & $11.56 \pm 0.36$ & $13.75 \pm 0.75$ & $12.76 \pm 0.68$ & $13.74 \pm 2.15$ \\
     & \textsc{pt-audio} & $11.58 \pm 0.12$ & $11.61 \pm 0.11$ & $11.68 \pm 0.34$ & $14.34 \pm 0.68$ & $13.84 \pm 0.68$ & $13.31 \pm 2.25$ \\
  & \textsc{pt-audio-fz} & $11.71 \pm 0.11$ & $11.71 \pm 0.10$ & $11.57 \pm 0.31$ & $12.80 \pm 0.71$ & $12.97 \pm 0.68$ & $11.11 \pm 2.47$ \\
& \textsc{pt-audio-eeg} & $11.67 \pm 0.12$ & $11.64 \pm 0.11$ & $11.69 \pm 0.34$ & $\bm{14.71 \pm 0.68}$ & $14.42 \pm 0.70$ & $\bm{14.46 \pm 2.42}$ \\
& \textsc{pt-audio-eeg-fz} & $\bm{11.45 \pm 0.10}$ & $\bm{11.46 \pm 0.10}$ & $\bm{11.43 \pm 0.32}$ & $14.56 \pm 0.70$ & $\bm{14.50 \pm 0.69}$ & $14.04 \pm 2.06$ \\

    \bottomrule
  \end{tabular}
  
\end{table*}

\section{Experiments}
\subsection{Dataset}
We conducted experiments on the N400 dataset \cite{n400}, which contains EEG signals that were recorded from 24 subjects, while each subject was listening to 440 sentences in English.
The speech samples are around 2 - 3 seconds long and contain 5 to 8 words.
The 128 channel EEG was recorded at a sampling rate of $512$ Hz, while the participants listened to the sentences.
Four subjects\footnote{subject \# 5, 10, 15, and 18.} were excluded owing to unreliable data, as suggested by \cite{n400}.
The test set consists of two subjects\footnote{subject \# 23 and 24.} and 40 sentences.
As a result, the train set contains 7,200 pairs\footnote{18 subjects with 400 sentences each.} and the remaining 1,600 pairs were selected for the test set, where 720 and 800 of them are unseen audio and subject, respectively, and 80 are unseen in both audio and subject.

The pre-processing pipeline for EEG is as follows: First, a notch filter at $60$ Hz was employed in order to remove the powerline noise. Then, bandpass filtering with frequency limits of low $0.5$ Hz and high $50$ Hz was applied in order to preserve the bands relevant to EEG spectral information, followed by eye blink removal using independent component analysis (ICA). The signals were then resampled to $256$ Hz.

All of the speech samples were down-sampled to $22,050$ Hz, to match the sampling rate of pre-trained speech models. For linear spectrograms, the following short-time Fourier transform (STFT) parameters were used: $1024$ for both FFT and window size, and $256$ for the hop size. For mel-spectrograms, the same parameters were utilized with $80$ mel-bands.

\subsection{Experimental Setup}
Five different training configurations
were considered as below:
\begin{itemize}
    \item \textsc{vanila}: all of the modules were trained from scratch, without any pre-trained parameters.
 \item \textsc{pt-audio}: only the speech module was initialized with pre-trained parameters for training.
 \item \textsc{pt-audio-fz}: the pre-trained parameters were used for the speech module, but frozen, that is, the speech module was not trained.
 \item \textsc{pt-audio-eeg}: both of the EEG module and the speech module were initialized with pre-trained parameters.
 \item \textsc{pt-audio-eeg-fz}: the EEG module and the speech module adopted the pre-trained parameters, but they were not trained.
\end{itemize}
For the pre-trained speech module,
\texttt{pretrained\_ljs.pth}\footnote{\url{https://github.com/jaywalnut310/vits}} was adopted.
In \textsc{pt-audio-eeg} and \textsc{pt-audio-eeg-fz}, the EEG module, that had been pre-trained for 300 epochs, was later frozen when combined with the speech module.
The compared baseline model is VLAAI \cite{vlaai}, which consists of a stack of convolution layers with skip connections.
The final layer of the baseline model was modified to generate 80 band mel-spectrograms. 
One Nvidia A40 GPU was utilized for each training configuration. All of the FESDE training configurations were trained for 100k iterations with the AdamW optimizer \cite{loshchilov2018decoupled}.
The baseline model was trained for 68 epochs.



\begin{figure*}[!ht]
  \centering
  \subfloat[MCD (dB) of each consonant (blue) and vowel (red) group.]{%
  \includegraphics[clip, width=0.98\textwidth,height=3.9cm]{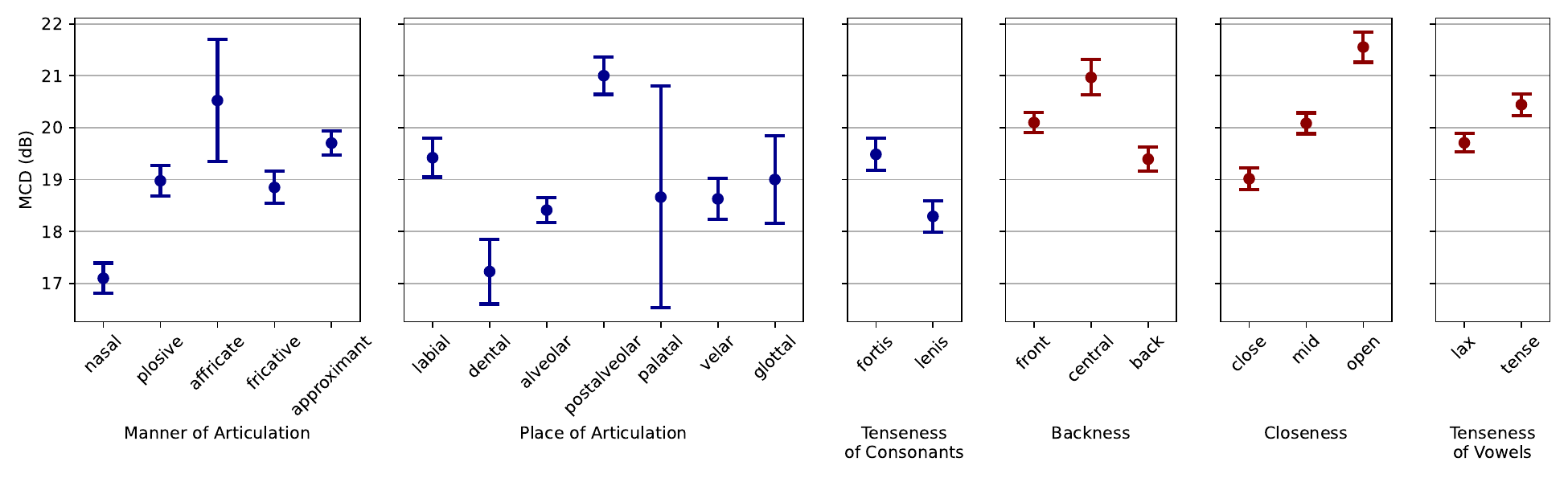}%
  }
  \vspace{-3mm}

  \subfloat[Mel-Corr (\%) of each consonant (blue) and vowel (red) group.]{%
  \includegraphics[clip, width=0.98\textwidth,height=3.9cm]{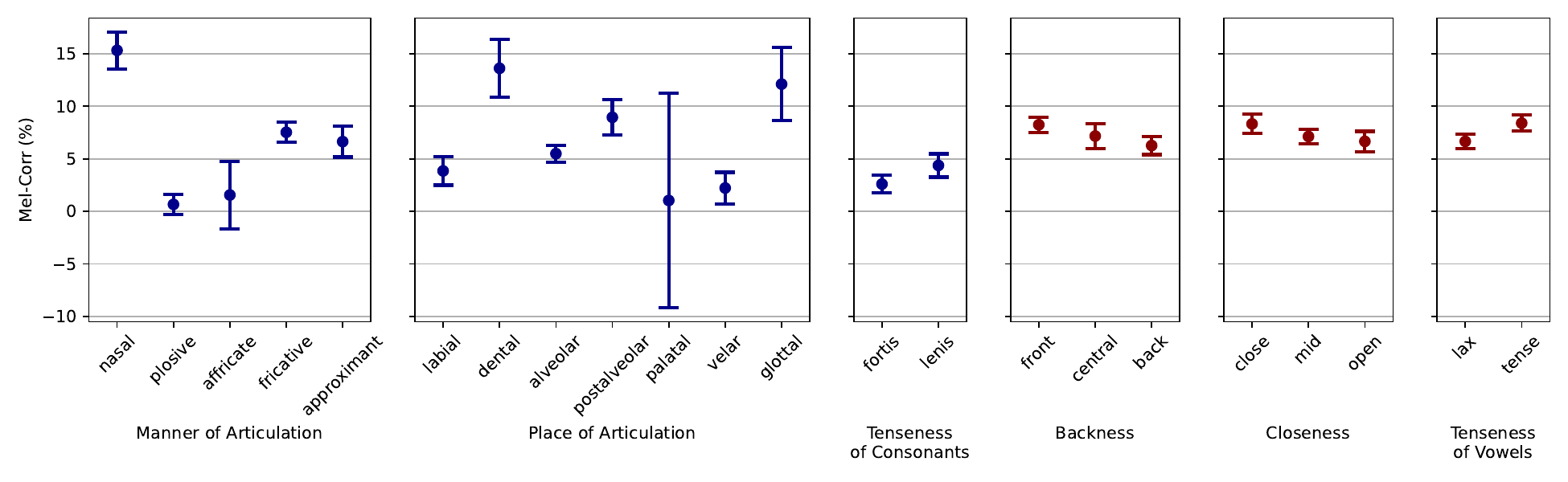}%
  }
  \caption{MCD (dB) and Mel-Corr (\%) of each phoneme group. The lower MCD and higher Mel-Corr indicate better performance. The consonants (blue) are clustered by three criteria: manner, place, and tenseness of articulation. The vowels (red) are clustered by its tongue position and tenseness.}
  \vspace{-3mm}
  \label{fig:phonemes}
\end{figure*}

\section{Results and Discussion}

\subsection{Evaluation}
The performance was measured in the following two objective metrics: mel-cepstral distortion (MCD) \cite{mcd} and mel-spectrogram correlation (Mel-Corr). MCD measures the distance between two mel-cepstral coefficients as Eq~(\ref{equation:mcd}):
\begin{align}
\vspace{-0.5cm}
\text{MCD} = \alpha\sqrt{\sum_{i=1}^{N_\text{MCC}}(\text{MCC}_i-\widehat{\text{MCC}_i})^2}
  \label{equation:mcd}
  \vspace{-0.5cm}
\end{align}
where $\alpha = \frac{10\sqrt{2}}{\ln10}$ and $N_{\text{MCC}}$ is the number of mel-cepstral coefficients.
Mel-Corr is calculated as the Pearson correlation coefficient between two mel-spectrograms. For convenience, the reported numbers are multiplied with 100.
Lower MCD and higher Mel-Corr values indicate better performance.

Table~\ref{tab:mcd-corr} shows the MCD and Mel-Corr values for the baseline model and the proposed five models. From the results in the table, the best performance was achieved when both of the EEG and speech modules are pre-trained separately (i.e., \textsc{pt-audio-eeg-fz}). Also, almost all of the proposed models perform better than the baseline model.

\subsection{Phoneme Analysis}
In order to analyze the effectiveness of proposed models in decoding speech at the level of phoneme, a fine-grained analysis was carried out by calculating some objective metrics.
This analysis will reveal which of the category of phonemes are difficult or easy to decode. The Montreal forced aligner (MFA) ver3 \cite{mfa} was used to acquire the timestamps of each phoneme. The pre-trained \texttt{english\_us\_arpa} configuration was used for both acoustic model and the dictionary. 
As MFA does not perform accurately for reconstructed speech, the timestamps acquired from the ground-truth speech are assumed to be identical to the reconstructed speech.

The consonants are grouped according to their manner of articulation, place of articulation, and tenseness of articulation. Similarly, the vowels are grouped based on position of the tongue and tenseness. As illustrated in Figures~\ref{fig:phonemes}, the consonants that are nasal, dental, or lenis tend to be relatively easily decoded. Also, an interesting tendency is observed wherein more closed vowels are easier to decode.


\section{Ablation Study}
\subsection{EEG Channels}
As recent studies \cite{eegsubset} suggest, we explore the utilization of only the parietal and temporal regions of EEG that are known to contain rich auditory information. 
Instead of an increase in performance, a drop in performance was observed, as shown in Table~\ref{tab:subchannel}, especially when the speech module was frozen during training.
This result may suggest that non-auditory parts of the brain may be involved in speech decoding, however, further investigation is necessary for a clearer explanation.
\begin{table}[t]
    \vspace{-0.1cm}
  \caption{Performance drop (\%) when only the parietal and temporal regions of EEG signals are used.}
    \vspace{-0.2cm}
  \label{tab:subchannel}
  \centering
  \begin{tabular}{lcc}
    \toprule
    \multicolumn{1}{l}{\textbf{Model}} & \multicolumn{2}{c}{\textbf{Performance Drop (\%)}} \\
     & \multicolumn{1}{c}{MCD} & \multicolumn{1}{c}{Mel-Corr}\\
    \toprule
\textsc{pt-audio-eeg} & $-0.13$ & $-5.99$ \\
 \textsc{pt-audio-eeg-fz}&  $-1.44$ & $-14.64$ \\
    \bottomrule
  \end{tabular}
      \vspace{-4mm}
\end{table}
\subsection{Text-Spotting}
Apart from evaluating the performance of speech reconstruction from EEG, we investigate whether our proposed EEG encoder captures textual information.
A binarized word-spotting experiment was conducted to validate whether our proposed EEG presentation can be used to identify certain words in a speech sample. 
The 30 most frequently occurring nouns among all sentences are chosen as the set of keywords, and the task is to detect whether any of these keywords exist in a speech. The pre-trained EEG module was utilized.
The results across different test conditions are presented in Table~\ref{tab:text_decoding}.
We demonstrate the feasibility of text detection from EEG signals, and it remains a challenging and open endeavor. 
\begin{table}[t!]
  \caption{Performance of predicting binarized word-spotting.
The performance score is measured in the unweighted F1 score.}
  \label{tab:text_decoding}
  \centering
  \begin{tabular}{ccc}
    \toprule
    \multicolumn{1}{c}{\textbf{unseen audio}} & \multicolumn{1}{c}{\textbf{unseen subject}} & \multicolumn{1}{c}{\textbf{unseen both}}\\
    \toprule
    $60.12$ & $56.74$ & $57.78$ \\
    \bottomrule
  \end{tabular}
\end{table}


\section{Conclusion}
In this work, we introduce FESDE, a framework to decode listened speech waveforms directly from EEG signals.
The proposed approach is faster and simpler by enabling single-step inference and it also outperforms the baseline model that involves an intermediate conversion to text representation.
We also explore the characteristics of phonemes that are easily or challenging to be decoded by the proposed framework.
Our approach is currently limited to listened speech, however, in the foreseeable future, we plan to extend our research scope to speech production tasks, such as imagined or phonated speech decoding.

\section{Acknowledgements}
This project was partially supported by a USC Annenberg Graduate Fellowship and also by the Defense Advanced Research Projects Agency (DARPA) under cooperative agreement No. N660012324006. The content of the information does not necessarily reflect the position or the policy of the Government, and no official endorsement should be inferred.


\bibliographystyle{IEEEtran}
\bibliography{mybib}

\end{document}